\documentclass[amssymb,aps]{revtex4}
\usepackage{graphicx}

\begin{document}

\title{Reply to ``Comment on ``Light-Front Schwinger Model at Finite
  Temperature''''} 

\author{Ashok Das and Xingxiang Zhou}
\affiliation{Department of Physics and Astronomy,
University of Rochester,
Rochester, NY 14627-0171, USA}
\bigskip

\begin{abstract}

In hep-th/0310278, Blankleider and Kvinikhidze criticize the
form of the thermal propagator used in \cite{das1} and  propose an alternate
thermal propagator for the fermions in the light-front Schwinger
model. We show that, within the standard light-front quantization used in
\cite{das1}, the thermal propagator for the fermions is unique as
presented in that paper.

\end{abstract}

\maketitle

In an earlier paper \cite{das1}, we studied various questions
associated with the light-front Schwinger model at finite temperature
where the theory was quantized on the standard 
light-front surface $\bar{x}^{0}= x^{0}+x^{1}=0$. (We refer the reader to
\cite{das1} for notations and conventions.) We had
argued in that paper that one of the components of the fermion field
does not thermalize and correspondingly 
had used the real time propagator (only the $++$ component of the
propagator and we will suppress the ``$i\epsilon$'' for simplicity)
\begin{eqnarray}
iS^{(T){\rm (DZ)}}_{+} (\bar{p}) & = & -
  \bar{p}_{1}\left(\frac{i}{-(2\bar{p}_{0}+\bar{p}_{1}) \bar{p}_{1}} -
  2\pi n_{F} (|\bar{p}_{0}|)
  \delta ((2\bar{p}_{0}+\bar{p}_{1})\bar{p}_{1})\right)\, ,\nonumber\\
iS^{(T){\rm (DZ)}}_{-} (\bar{p}) & = & \frac{i}{-\bar{p}_{1}}\,
  ,\label{1} 
\end{eqnarray} 
where $n_{F} (|\bar{p}_{0}|)$ represents the Fermi-Dirac distribution
function. This propagator was obtained from the structure of the
fermion Lagrangian density of the light-front Schwinger model and was
not derived from
Eq. (33) in \cite{das1} which describes the propagator for a massive
fermion in higher dimensions. (The authors in \cite{blankleider} seem to
suggest that our propagator was derived from an erroneous limit of that
expression.) We had, however, indicated that Eq. (\ref{1}) can be obtained
from Eq. (33) in \cite{das1} in a limiting manner. One of the results
found in that paper showed that 
the off-shell  thermal $n$-point photon amplitudes in this theory do not
coincide with the ones calculated in
the conventionally quantized (equal-time) Schwinger model \cite{das2}
and we had traced the origin of the difference to the fact that one of
the fermion components in the light-front Schwinger model is
nondynamical in the quantization used and as a result does not
thermalize, while both the
fermion components in the conventionally quantized theory are
dynamical and do thermalize. {\em We note, on
the other hand, that all the thermal corrections to the $n$-point photon
amplitudes vanish at zero
temperature in both the quantizations and the $n$-point photon amplitudes  
do coincide.} 

In \cite{blankleider} the authors comment that the
difference found in \cite{das1} has its origin in the use of an
erroneously simplified thermal fermion propagator and suggest that the
proper thermal propagator for the
fermions that should have been 
used in \cite{das1} is of the form (only the $++$ component)
\begin{eqnarray}
iS^{(T) {\rm (BK)}}_{+} (\bar{p}) & = & -
  \bar{p}_{1}\left(\frac{i}{-(2\bar{p}_{0}+\bar{p}_{1}) \bar{p}_{1}} -
  2\pi n_{F} (|\bar{p}_{0}|)  \delta
  ((2\bar{p}_{0}+\bar{p}_{1})\bar{p}_{1})\right)\, ,\nonumber\\ 
iS^{(T) {\rm (BK)}}_{-} (\bar{p}) & = &
  (2\bar{p}_{0}+\bar{p}_{1})\left(\frac{i}{-(2\bar{p}_{0}+\bar{p}_{1})
  \bar{p}_{1}} - 2\pi n_{F}
  (|\bar{p}_{0}|) \delta
  ((2\bar{p}_{0}+\bar{p}_{1})\bar{p}_{1})\right)\, .\label{2}
\end{eqnarray}
They obtain this propagator from Eq. (33) in \cite{das1} by 
setting $m=0$ (and restricting to $1+1$ dimensions). The difference
between (\ref{1}) and (\ref{2}) lies in the thermal part of the propagator
$iS^{(T)}_{-} (\bar{p})$. Namely, their contention is that both
components of the fermion field in the light-front 
Schwinger model should thermalize, even though one of them is
nondynamical. In this case, of course, one should not expect any
difference from the results of the conventionally quantized
theory. The basic issue,
therefore, is whether the $\psi_{-}$ component in the light-front
Schwinger model thermalizes in the quantization used in \cite{das1}.

Given the quantization conditions in a theory, the
propagators are, of course, uniquely determined as vacuum expectation
values  of time ordered products of fields. Therefore, it is not
entirely clear from \cite{blankleider},
whether the authors find the
fermion propagator in \cite{das1} to be incorrect within the
quantization used or whether their objection is addressed to the
quantization
used in that paper. We will try to address both these issues in the
following.
 
First, let us discuss the form of the propagator within the
quantization used in \cite{das1}. There are various ways to see, both
in the  imaginary time and the real
time formalisms, that the $\psi_{-}$ component in the light-front
Schwinger model {\it does not
  thermalize} in the standard light-front quantization used in
\cite{das1}. We briefly discuss the imaginary time
formalism before going into the real time formalism. We note that the
quadratic part of the fermion
Lagrangian density (which is relevant for a discussion of the
propagator) for the light-front Schwinger model has the form
\begin{equation}
{\cal L} = i \psi_{+}^{\dagger} (2\bar{\partial}_{0} +
\bar{\partial}_{1}) \psi_{+} - i \psi_{-}^{\dagger}
\bar{\partial}_{1}\psi_{-}\, .\label{3}
\end{equation}
Here $\psi_{+},\psi_{-}$ represent the two chiral components of the
theory. The zero
temperature propagators of the theory in (\ref{3}) have the
simple forms
\begin{equation}
iS^{(0)}_{+} (\bar{p}) = - \frac{i\bar{p}_{1}}{-(2\bar{p}_{0}+\bar{p}_{1})
  \bar{p}_{1}} = - \frac{i}{-(2\bar{p}_{0}+\bar{p}_{1})},\qquad
  iS^{(0)}_{-} 
  (\bar{p}) = \frac{i
  (2\bar{p}_{0}+\bar{p}_{1})}{-(2\bar{p}_{0}+\bar{p}_{1})\bar{p}_{1}}
  = \frac{i}{-\bar{p}_{1}}\, .\label{4}
\end{equation}
In the imaginary time formalism in light-front theories within the
quantization used in \cite{das1}, one obtains
the thermal propagators simply by letting \cite{das3,weldon}
\begin{equation}
\bar{p}_{0} \rightarrow (2n+1)i\pi T\, ,\label{5}
\end{equation}
where $T$ denotes temperature. This introduces a temperature
dependence to $iS^{(T)}_{+} (\bar{p})$ in (\ref{4}), while
$iS^{(T)}_{-} 
(\bar{p})$  remains temperature independent since it
does not depend on 
$\bar{p}_{0}$. This is probably the most direct way to see
that the component $\psi_{-}$ does not thermalize in the standard
light-front quantization.

Let us next analyze the propagator in the real time formalism. This is
best done in the operatorial formalism of thermofield dynamics
\cite{umezawa,das4}. We note that a Hamiltonian analysis of the theory
in (\ref{3}) shows that, when quantized on the surface $\bar{x}^{0}=0$,
the only nontrivial anti-commutation relation has the form
\begin{equation}
\left\{\psi_{+} (\bar{x}), \psi^{\dagger}_{+}
(\bar{y})\right\}_{\bar{x}^{0}=\bar{y}^{0}} =  P^{+} \delta
(\bar{x}^{1}-\bar{y}^{1})\,,\label{6}
\end{equation}
where $P^{+}$ represents the projection operator for the positive
chirality spinors.  
Since the fermion field $\psi_{-}$ satisfies trivial anti-commutation
relations, it follows in particular that
\begin{equation}
\left\{\psi_{-} (\bar{x}), H\right\} = 0\,,\label{7}
\end{equation}
namely, the $\psi_{-}$ component has no time evolution. As a result,
the propagator for the $\psi_{-}$ field has the form
\begin{equation}
iS^{(0)}_{-} (\bar{x}-\bar{y}) = \langle 0|{\rm T} \left(\psi_{-} (\bar{x})
\psi^{\dagger}_{-} (\bar{y})\right)|0\rangle = \langle 0|\psi_{-}
(\bar{x}) \psi^{\dagger}_{-} (\bar{y})|0\rangle =
F(\bar{x}^{1}-\bar{y}^{1})\,,\label{8}
\end{equation}
which is consistent with the form of the zero temperature propagator
$iS_{-}^{(0)}$ in (\ref{4}).

In going to finite temperature, in thermofield dynamics, one doubles
the degrees of freedom (with tilde fields) and obtains a thermal
vacuum through a Bogoliubov transformation of the form
\begin{equation}
|0(\beta)\rangle = U(\theta) |0\rangle \otimes |\tilde{0}\rangle\,
 ,\label{9}
\end{equation}
where $\beta$ represents the inverse temperature in units of the
Boltzmann constant. The formally unitary transformation involves
both the physical and the tilde fields and has the
form  
\begin{equation}
U (\theta) = e^{-iQ(\theta)}\, ,\label{10}
\end{equation}
with the parameter $\theta$ related to the fermion distribution
function \cite{umezawa,das4}. The finite temperature propagator for
the $\psi_{-}$  field is then defined in the standard manner as
\begin{equation}
iS^{(\beta)}_{-} (\bar{x}-\bar{y}) = \langle 0(\beta)|{\rm
  T}\left(\psi_{-} (\bar{x}) 
\psi^{\dagger}_{-} (\bar{y})\right)|0(\beta)\rangle\,.\label{11}
\end{equation}
From  (\ref{8})-(\ref{11}) as well as the fact that $\psi_{-}$
satisfies trivial anti-commutation relations in the standard
light-front quantization, it follows that
\begin{eqnarray}
iS^{(\beta)}_{-} (\bar{x}-\bar{y}) & = & \langle \tilde{0}|\otimes
\langle 0|\left(e^{iQ(\theta)}\psi_{-} (\bar{x})\psi^{\dagger}_{-}
(\bar{y}) e^{-i Q(\theta)}\right)|0\rangle \otimes |\tilde{0}\rangle =
\langle \tilde{0}|\otimes \langle 0|\left(\psi_{-}
(\bar{x})\psi^{\dagger}_{-} (\bar{y})\right)|0\rangle \otimes
|\tilde{0}\rangle \nonumber\\
 & = & \langle 0|\psi_{-} (\bar{x}) \psi^{\dagger}_{-}
(\bar{y})|0\rangle = iS^{(0)}_{-} (\bar{x}-\bar{y}) = F
(\bar{x}^{1}-\bar{y}^{1})\, .\label{12}
\end{eqnarray}
This demonstrates clearly that within the standard light-front
quantization used in \cite{das1}, the fermion field $\psi_{-}$ does
not thermalize and that the unique finite temperature propagator coincides
with that at zero temperature which is the the form used in \cite{das1}.

As we had indicated in \cite{das1}, this form of the propagator can
also be obtained from a limit of  Eq. (33) (a massive propagator) of
that paper. Essentially, this involves looking at the vanishing mass
limit of a delta function of the form
$(2\bar{p}_{0}+\bar{p}_{1})\delta
((2\bar{p}_{0}+\bar{p}_{1})\bar{p}_{1} + m^{2})$. If $m=0$, there are
two roots for the vanishing of the delta function. Keeping both the
roots leads to the propagator in (\ref{2}) which, however, would not be 
compatible with (\ref{12}) and would lead to a nontrivial time
dependence  (in the coordinate space). Therefore, naively setting $m=0$ in
Eq. (33) of \cite{das1} would not lead to the proper propagator within
the quantization being discussed. The propagator in (\ref{12}) (and,
therefore, (\ref{1})), on the other hand, can be obtained from a
massive theory (Eq. (33)
of \cite{das1}) only if a particular limiting value is chosen (namely,
$|\bar{p}_{0}|,|\bar{p}_{1}|\gg m\rightarrow 0$)  which selects out 
only the root $(2\bar{p}_{0}+\bar{p}_{1})=0$ of the delta function. As
is also noted in \cite{blankleider}, the massless limit
in light-front theories is subtle; therefore, when necessary one must go back
to the basic definitions, as we have just done for the
propagator (and as was also done in \cite{das1}). 

The authors of \cite{blankleider} have also argued how the $\psi_{-}$ field
can become dynamical in the non-standard light-front quantization due
to McCartor \cite{mccartor} which involves quantizing the $\psi_{+}$
field on the conventional surface $x^{+} = x^{0}+x^{1}=0$ while
quantizing the $\psi_{-}$ component on the surface $x^{-} =
x^{0}-x^{1}=0$. This brings us to the question of whether their
objection is really to the quantization used in \cite{das1}. It is worth
recognizing that a given quantization
defines a unique quantum theory and different quantizations do not
yield equivalent quantum theories in general. As McCartor
himself has pointed out \cite{mccartor1}, his non-standard quantization
leads to a vanishing fermion condensate (in the infinite volume limit)
which is in disagreement with all the other calculations. The theory
quantized on $\bar{x}^{0}=x^{0}+x^{1}=0$, on the other hand, does lead
to the condensate \cite{prem} (even at finite temperature \cite{das1})
which agrees with the results of equal-time quantization. Therefore,  
it is not clear {\em a priori} which of the two theories should be
called the light-front Schwinger model (if that is the objection being
raised by the authors in \cite{blankleider}). It is, of course, an
interesting question to see if McCartor's alternative quantization (or a
generalization of it) does allow a
statistical description (We remind the reader that the conventional
light-front quantization does not.) and if so whether it leads to the
propagator in Eq. (\ref{2}) at finite temperature. Even if it does,
that would 
not be the appropriate propagator to use in a calculation involving
the standard light-front quantization such as in \cite{das1}.
As we have argued above the propagator used in \cite{das1} is the unique
propagator within that quantization and leads to the result that {\em
  at zero temperature the $n$-point photon amplitudes agree with the
  calculations using equal-time relations, while at finite
  temperature, the results are different}. 

\vspace{1cm}

\noindent{\bf Acknowledgment:}

This work was supported in part by US DOE Grant number DE-FG 02-91ER40685.

\end{document}